\documentclass[a4paper]{jpconf}
\usepackage{graphicx}
\begin{document}
\title{Ultracold and dense samples of ground-state molecules in lattice potentials}

\author{Hanns-Christoph N\"agerl, Manfred J. Mark, Elmar Haller, Mattias Gustavsson, Russell Hart, and Johann G. Danzl}

\address{Institut f\"{u}r Experimentalphysik und Forschungszentrum f\"{u}r Quantenphysik, Universit\"{a}t Innsbruck, 6020 Innsbruck, Austria}

\ead{christoph.naegerl@uibk.ac.at}

\begin{abstract}
We produce an ultracold and dense sample of rovibronic ground state Cs$_2$ molecules close to the regime of quantum degeneracy, in a single hyperfine level, in the presence of an optical lattice. The molecules are individually trapped, in the motional ground state of an optical lattice well, with a lifetime of 8 s. For preparation, we start with a zero-temperature atomic Mott-insulator state with optimized double-site occupancy and efficiently associate weakly-bound dimer molecules on a Feshbach resonance. Despite extremely weak Franck-Condon wavefunction overlap, the molecules are subsequently transferred with $>$50\% efficiency to the rovibronic ground state by a stimulated four-photon process. Our results present a crucial step towards the generation of Bose-Einstein condensates of ground-state molecules and, when suitably generalized to polar heteronuclear molecules such as RbCs, the realization of dipolar many-body quantum-gas phases in periodic potentials.
\end{abstract}

\section{Introduction}
Over the last 15 years there has been tremendous progress in controlling atomic gaseous quantum matter at ultralow temperatures in laboratory experiments. At the same time there have been spectacular advances in the theoretical understanding of ultracold matter. For example, quantum degenerate atomic samples have been loaded into optical lattice potentials, allowing the realization of strongly-correlated many-body systems and enabling the direct observation of quantum phase transitions with essentially full control over the parameter space [1--4]\nocite{Jaksch1998cba,Greiner2002qpt,Bloch2008mbp,Haller2010pqp}. Also, quantum gas mixtures have been at the focus of experimental and theoretical interest. For example, fermionic spin mixtures have enabled ground-breaking experiments on the BEC-BCS crossover (for an overview, see Ref.~\cite{Varenna2008ufg,Giorgini2008tou}). Bose-Bose and Fermi-Bose quantum gas mixtures have recently attracted further widespread attention as a starting point for efficient creation of high phase-space density samples of weakly-bound molecules \cite{Regal2003cum,Herbig2003poa} created on Feshbach resonances and, after molecular state transfer, the creation of ultracold and dense samples of deeply bound ground-state molecules [9--11]\nocite{Danzl2008qgo,Ni2008ahp,Danzl2010auh}. Such molecular samples promise myriad possibilities for new directions of research in fields such as ultracold chemistry, quantum information science, precision metrology, Bose-Einstein condensation, and quantum many-body physics
[12--17]\nocite{Krems2005mna,Krems2008ccc,Demille2002qcw,DeMille2008est,Zelevinsky2008pto,Goral2002qpo}
(for recent overview articles on the status of the field and on possible experiments, see Refs.~[18--23]\nocite{Baranov2008tpi,Pupillo2008cmp,Lahaye2009tpo,Carr2009cau,Friedrich2009wac,Dulieu2009tfa}).
Ultracold samples of homonuclear molecules such as Cs$_2$ would allow novel studies of few-body collisional physics [24--28]\nocite{Chin2005oof,Kraemer2006efe,Staanum2006eio,Zahzam2006amc,Idziaszek2010sqm}, would enable Bose-Einstein condensation of ground state molecules \cite{Jaksch2002coa}, and would open up new terrain for high resolution spectroscopy \cite{Flambaum2007est,DeMille2008est,Zelevinsky2008pto}. Heteronuclear molecules such as KRb and RbCs, with sizable permanent electric dipole moments of the order of 1 Debye in their rovibronic ground state \cite{Kotochigova2005air,Deiglmayr2008cos}, feature long-range and anisotropic dipole-dipole interactions. Therefore, ultracold samples of polar molecules open up completely new avenues for research. For example, at ultralow temperatures, scattering and reaction dynamics are expected to be dominated by dipolar effects [28, 33--39]\nocite{Micheli2007cpm,Micheli2010urf,Ticknor2010qtd,Quemener2010sdo,Quemener2010efs,Idziaszek2010urc,Idziaszek2010sqm,Li2010uic}. In fact, recent experiments on KRb have reached this regime \cite{Ni2010dco}. A large variety of novel quantum phases has been proposed for bosonic dipolar gases that are confined to two- or three-dimensional lattice potentials
[17, 41--51]\nocite{Goral2002qpo,Barnett2006qmw,Buechler2007sc2,Menotti2007mso,Yi2007nqp,Danashita2009sos,Cooper2009sts,Capogrosso2010qpo,Pollet2010spw,Li2010cmo,Potter2010sad,Pikovski2010isi}. In particular, strongly coupled ladder systems of strongly-interacting one-dimensional (1D) gases should become accessible [52--56]\nocite{DallaTorre2006hoi,Kollath2008dbi,Huang2009qpd,Klawunn2010bfm,Dalmonte2010odq}, allowing us to greatly extend recent work in our group on strongly interacting 1D systems \cite{Haller2009roa,Haller2010pqp} to the case of dipolar systems.

Here, we review and extend our work \cite{Danzl2010auh} on producing ultracold and dense samples of ground state Cs$_2$ molecules individually trapped at the sites of an optical lattice. First, we create ultracold samples of weakly bound Feshbach molecules out of an atomic Mott-insulator state by means of a Feshbach resonance \cite{Kohler2006poc,Chin2010fri}. While maintaining ultralow temperatures and state fidelity, the molecules are then transferred into the ro-vibrational ground state by a stimulated four-photon transition using the STIRAP technique \cite{Bergmann1998cpt}. Recently, the STIRAP technique has been successfully implemented with high efficiencies on ultracold and dense samples of Feshbach molecules for transfer into the rovibrational ground state of the triplet $^3\Sigma$-potential for Rb$_2$ molecules \cite{Lang2008utm}, for transfer into deeply bound rovibrational levels of the singlet $^1\Sigma$-potential for Cs$_2$ molecules \cite{Danzl2008qgo} and also, in a four-photon process, into the rovibronic ground state \cite{Danzl2010auh} of this molecule, and for transfer into the rovibronic ground state for polar KRb molecules \cite{Ni2008ahp}.

\section{Experimental procedure}

\begin{figure}[h]
\includegraphics{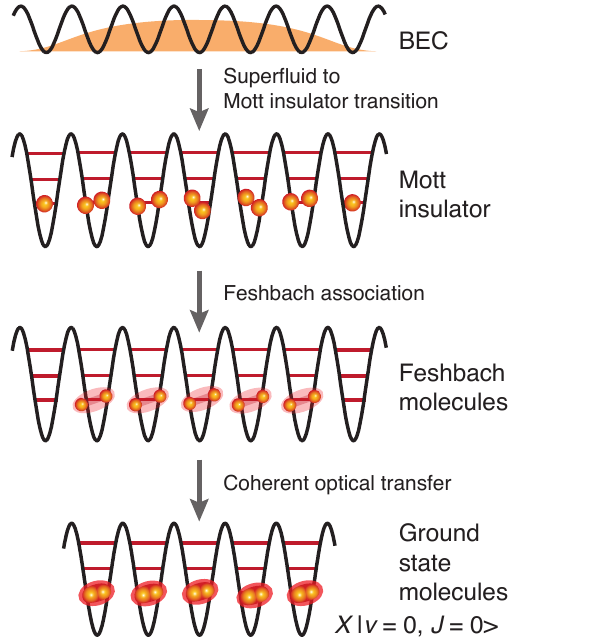}\hspace{2pc}
\begin{minipage}[b]{20pc}\caption{\label{ICAPfig1}{\bf Molecular quantum gas preparation procedure:} A BEC of Cs atoms is loaded into an optical lattice. The superfluid-to-Mott-insulator quantum phase transition is driven under conditions that maximize the number of doubly occupied lattice sites. Feshbach association creates weakly bound molecules with each molecule individually trapped in the lowest vibrational level of its respective lattice site. Atoms remaining at singly occupied sites are cleared out. The molecules are then coherently transferred by STIRAP to a specific hyperfine quantum state of the rovibronic ground state $X^1\Sigma_g^+ |v=0,J=0\!\!>$. They remain trapped in the lowest lattice vibrational level.}
\end{minipage}
\end{figure}

\begin{figure}[h]
\begin{center}
\includegraphics[width=32pc]{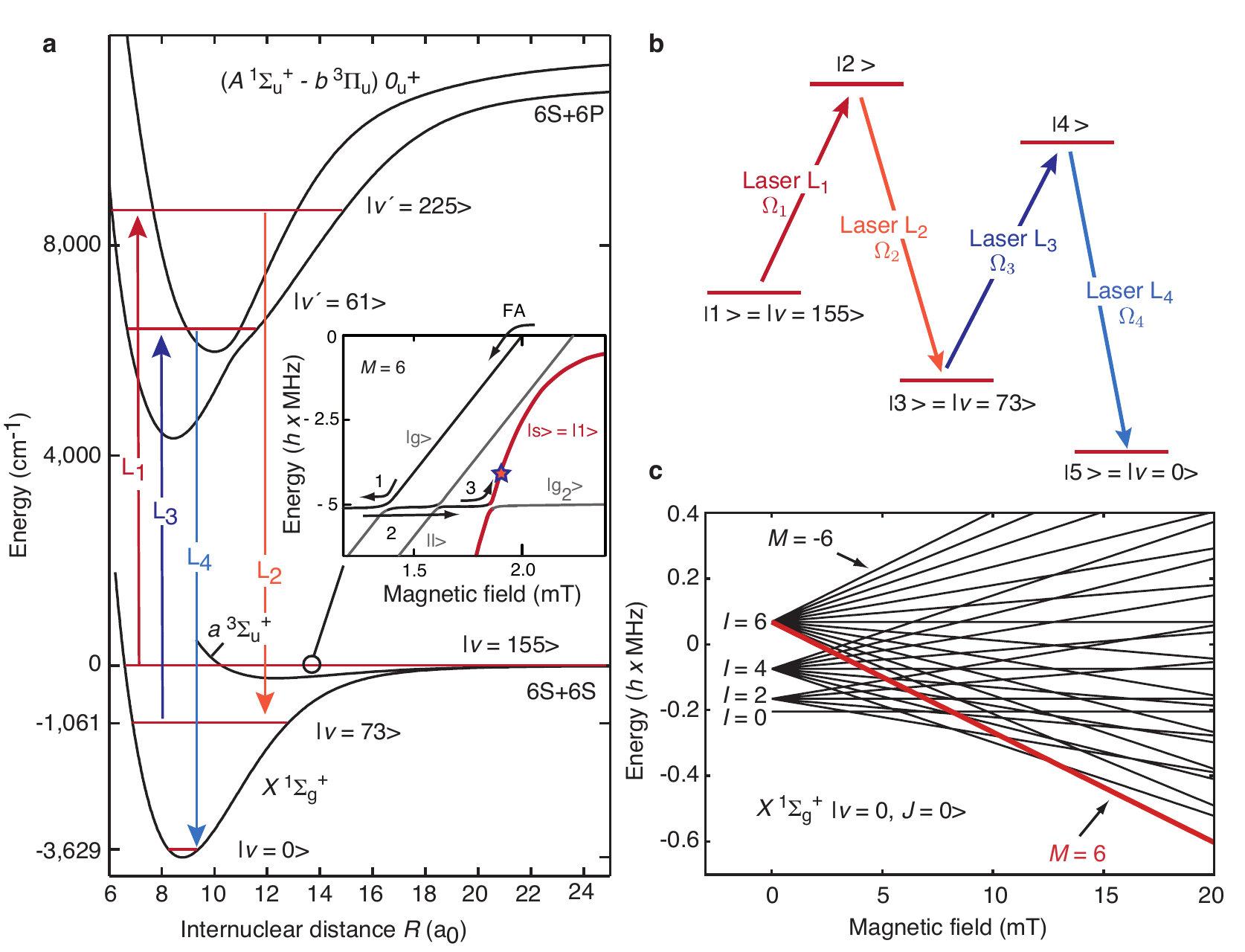}
\end{center}
\caption{\label{ICAPfig2}{\bf Molecular potentials and level schemes for transfer to the rovibronic ground state: a,} Molecular potentials. The initial Feshbach level (not resolved near threshold) of mixed $(a^3\Sigma_u^+ - X^1\Sigma_g^+)$ character is linked to the rovibronic ground state by 4 laser transitions via two excited state levels belonging to the $(A^1\Sigma_u^+-b^3\Pi_u)0_u^+$ coupled system and one ground state intermediate level $|v=73\!\!>$. The transition wavelengths for lasers L$_i$, where $i=1,2,3,4$, are near 1126 nm, 1006 nm, 1351 nm, and 1003 nm, respectively. Inset: Zeeman diagram for the molecular levels near threshold. Molecules are created by Feshbach association (FA) and the molecular population is transferred to the desired starting level for STIRAP $|s\!\!>=|1\!\!>$ ($\star$) via avoided level crossings by slow (1,3) and fast (2) magnetic field ramps. All weakly bound levels are characterized by $M=6$. {\bf b,} Schematic 5-level scheme linking the initial weakly bound level $|1\!\!>$ to the rovibronic ground state $|5\!\!>$ by lasers L$_i$ with Rabi frequencies $\Omega_i$. {\bf c,} Zeeman diagram showing the calculated hyperfine structure of the rovibronic ground state $|5\!\!>=|v=0,J=0\!\!>$. Coherent optical transfer selectively populates the level $|I=6, M_I=6\!\!>$ indicated in red. This is the lowest energy absolute ground state at magnetic fields above $\sim$ 13 mT. There are no avoided crossings between hyperfine sublevels. Calculation taken from Ref. \cite{Danzl2010auh}.}
\end{figure}

Our molecular quantum-gas preparation procedure is summarized in Fig.\,\ref{ICAPfig1}. We first load a pure Bose-Einstein condensate (BEC) of about $10^5$ Cs atoms \cite{Kraemer2004opo} into a three-dimensional, far detuned optical lattice. By raising the lattice depth and adjusting the external trap confinement we drive the superfluid-to-Mott-insulator (SF-MI) phase transition \cite{Bloch2008mbp} to localize pairs of atoms at individual lattice sites. The atomic number distribution in the Mott-insulator state is inhomogeneous as a result of the external harmonic confinement. Shells with a constant number of precisely $n$ atoms per lattice site, where $n=1,2,3\ldots$, are separated by narrow superfluid regions \cite{Bloch2008mbp}. By appropriately increasing the strength of the external confinement and hence compressing the sample  during lattice loading, we aim to create a ($n\!\!=\!\!2$)-Mott shell with maximum extent in the central region of the lattice in order to obtain the highest number of lattice sites at which there are precisely two atoms. With up to 45(2)\% of the atoms at doubly-occupied lattice sites we come close to the theoretical limit of 53\% given the parabolic density profile of the BEC \cite{Volz2006pqs}. The atom pairs reside in the motional ground state at each well and are then associated \cite{Thalhammer2006llf} with 94(1)\% probability \cite{Danzl2010auh} to Cs$_2$ Feshbach molecules, which are subsequently transferred via a series of avoided level crossings to the weakly-bound level $|1\!\!>$, the starting level for the optical transfer \cite{Mark2007sou,Danzl2008qgo,Danzl2009dbu}, see Fig.\,\ref{ICAPfig2}.
We typically set the depth of the optical lattice potential to 20 $E_R$ for atoms, corresponding to 80 $\tilde{E}_R$ for Feshbach molecules with double the atomic polarizability and double the mass, where $E_R$ and $\tilde{E}_R$ are the atomic and molecular photon recoil energies, respectively.
Atoms at singly-occupied sites are removed by a combination of microwave and optical excitation \cite{Thalhammer2006llf}. We now have a pure molecular sample with a high occupation of about 85(3)\% in the central region of the lattice.
Each molecule is in the motional ground state of its respective lattice well and perfectly shielded from collisional loss.

The molecular population is transferred to the rovibronic ground state by coherent optical transfer using the Stimulated Raman Adiabatic Passage (STIRAP) technique. STIRAP allows highly efficient population transfer between molecular levels with exquisite state selectivity, since there are no spontaneous processes involved \cite{Bergmann1998cpt}.
The molecular population is rotated to the final molecular level via a dark state which has no contribution from lossy electronically excited levels. STIRAP is effected by a sequence of overlapping laser pulses, which adiabatically changes the character of the dark state from the initially populated molecular level to the target level.
To satisfy adiabaticity, the product of the STIRAP transfer time and the Rabi frequencies for the STIRAP transitions needs to be sufficiently high. For transfer from the extremely weakly bound molecular levels, populated by Feshbach association, to the rovibronic ground state, this requirement poses a considerable challenge. On the one hand, there is a large mismatch in internuclear separation that has to be bridged.
On the other hand, the lasers involved in molecular state transfer need to be phase coherent with respect to each other on the timescale of the STIRAP process. Long STIRAP transfer times, while beneficial for adiabaticity, pose stringent requirements on the relative phase stability of the transfer lasers operating at widely different wavelengths.

We have identified a transfer scheme that allows us to efficiently transfer the molecular population to the rovibronic ground state. A total of four molecular transitions, driven by lasers L$_1$ \ldots L$_4$, couple the initial weakly bound level $|1\!\!>$ with the rovibronic ground state $|5\!\!>$ via two intermediate electronically excited state levels $|2\!\!>, |4\!\!>$ and one intermediate ground state level $|3\!\!>$, as illustrated in Fig.\,\ref{ICAPfig2}a and \ref{ICAPfig2}b. The four-photon (4p) scheme is preferred over a two-photon scheme to overcome otherwise prohibitively low Franck-Condon (FC) factors. When compared to heteronuclear molecules, homonuclear molecules, such as Cs$_2$, feature an additional symmetry with respect to exchange of the identical nuclei, the so-called $gerade/ungerade$ $(g/u)$ symmetry, which imposes additional selection rules for optical transitions and hence restricts the excited state potentials that can be employed in the optical transfer. In addition, a resonant term in the interaction leads to a $1/R^{3}$ scaling of the excited state potentials as opposed to the $1/R^{6}$ scaling in the ground state, implying that ground and excited state potentials do not line up favorably for transfer from weakly to deeply bound levels. While adding some complexity to the system, the four-photon scheme allows acceptable, albeit low, FC factors for coherent state transfer. Specifically, both excited state intermediate levels $|2\!\!>$ and $|4\!\!>$belong to the mixed singlet/triplet $(A^1\Sigma_u^+-b^3\Pi_u)0_u^+$ coupled system, as illustrated in Fig.\,\ref{ICAPfig2}a. The intermediate ground state level $|3\!\!>$ is bound by $\sim hc \times 1060$ cm$^{-1}$ or $\sim h \times 32$ THz, where $h$ is Planck's constant and $c$ is the speed of light, and has vibrational quantum number $v=73$. We choose either $J=0$ or $J=2$, where $J$ is the rotational quantum number.
In the first transition, the singlet $X^1\Sigma_g^+$ component of the initial Feshbach level enables coupling to the $0_u^+$ system. In contrast to heteronuclear molecules, where singlet/triplet mixing in the excited state can be favourably exploited \cite{Stwalley2004eco} in combination with excitation from the dominant $a^3\Sigma^+$ component of Feshbach levels, a transition from the $a^3\Sigma_u^+$ state to $0_u^+$ levels is forbidden here.
We have performed ultra-high-resolution molecular spectroscopy to identify the previously unknown excited state intermediate molecular levels in the four-photon scheme \cite{Danzl2009pms,Mark2009drf}. These have highly mixed singlet/triplet character. Interestingly, for the second two-photon transition, the excited state intermediate level $|4\!\!>$ is primarily of triplet character, giving a more balanced distribution of FC factors for lasers L$_3$ and L$_4$ than neighboring levels of primarily singlet character.

Our experimental procedure ensures that only a single quantum state is populated in $X |v=0,J=0\!\!>$. The rovibrational ground state of the $^1\Sigma$ electronic ground state in alkali dimers comprises a series of hyperfine sublevels separated by a very small energy at low magnetic fields \cite{Aldegunde2008hel,Aldegunde2009hel}. In Cs$_2$, there are 28 hyperfine sublevels in $|v=0,J=0\!\!>$ separated by $\sim h \times 270$ kHz at zero field \cite{Danzl2010auh}, as illustrated in Fig.\,\ref{ICAPfig2}c. These correspond to the allowed values for the total molecular nuclear spin $I=0,2,4,6$ with the $(2I+1)$ projections on the magnetic field. In homonuclear dimers there are no avoided crossings between hyperfine sublevels for $J=0$ levels \cite{Aldegunde2009hel}.
For comparison, the inset in Fig.\,\ref{ICAPfig2}a shows the weakly bound molecular levels near the dissociation threshold. We prepare the atomic BEC in the lowest atomic hyperfine sublevel $|F_a=3, m_{F_a}=3\!\!>$, where $F_a$ and $m_{F_a}$ are the total atomic spin and its projection on the magnetic field.
Feshbach association and transfer between weakly bound levels via avoided crossings conserve \cite{Chin2010fri} the total projection quantum number $M=m_{Fa_1}+m_{Fa_2}=6$.
In the coherent optical transfer, we exploit the selection rule $\Delta M=0$ for linear polarization parallel to the quantization axis, such that the rovibronic ground state molecules inherit the $M=6$ total angular momentum projection. From the Zeeman diagram in Fig.\,\ref{ICAPfig2}c it is immediately clear that there is only a single $M=6$ quantum level in the rovibronic ground state, namely the $|I=6,M_I=6\!\!>$ level, and hence a single, well-defined quantum state is selectively populated by our optical state transfer. Importantly, the $|I=6,M_I=6\!\!>$ quantum state is an energetically excited hyperfine sublevel at low magnetic fields at which the STIRAP transfer is carried out, but it constitutes the lowest hyperfine sublevel and hence the lowest energy absolute ground state of the Cs dimer at magnetic fields above $\sim 13$ mT.

To ensure phase coherence of the STIRAP lasers on the timescale of the optical transfer, the lasers are locked to narrow linewidth optical resonators for short term stability and referenced to an optical frequency comb for long-term stability and reproducibility.

\section{Results}

There are two possibilities to transfer the molecular population from $|1\!\!>$ to $|5\!\!>$. In sequential STIRAP (s-STIRAP), the molecular population is first transferred to the intermediate ground state level $|3\!\!>$ in a conventional two-photon process and then in a second two-photon step to $|v=0,J=0\!\!>$. In contrast, 4-photon STIRAP (4p-STIRAP) transfers the molecular population directly into $|v=0,J=0\!\!>$ via a 4p-dark state \cite{Malinovsky1997sar}, which has no contribution from the two excited state intermediate levels $|2\!\!>$ and $|4\!\!>$. A recent analysis in the context of ultracold molecules can be found in Ref. \cite{Kuznetsova2008fod}.

\begin{figure}[h]
\begin{center}
\includegraphics[width=32pc]{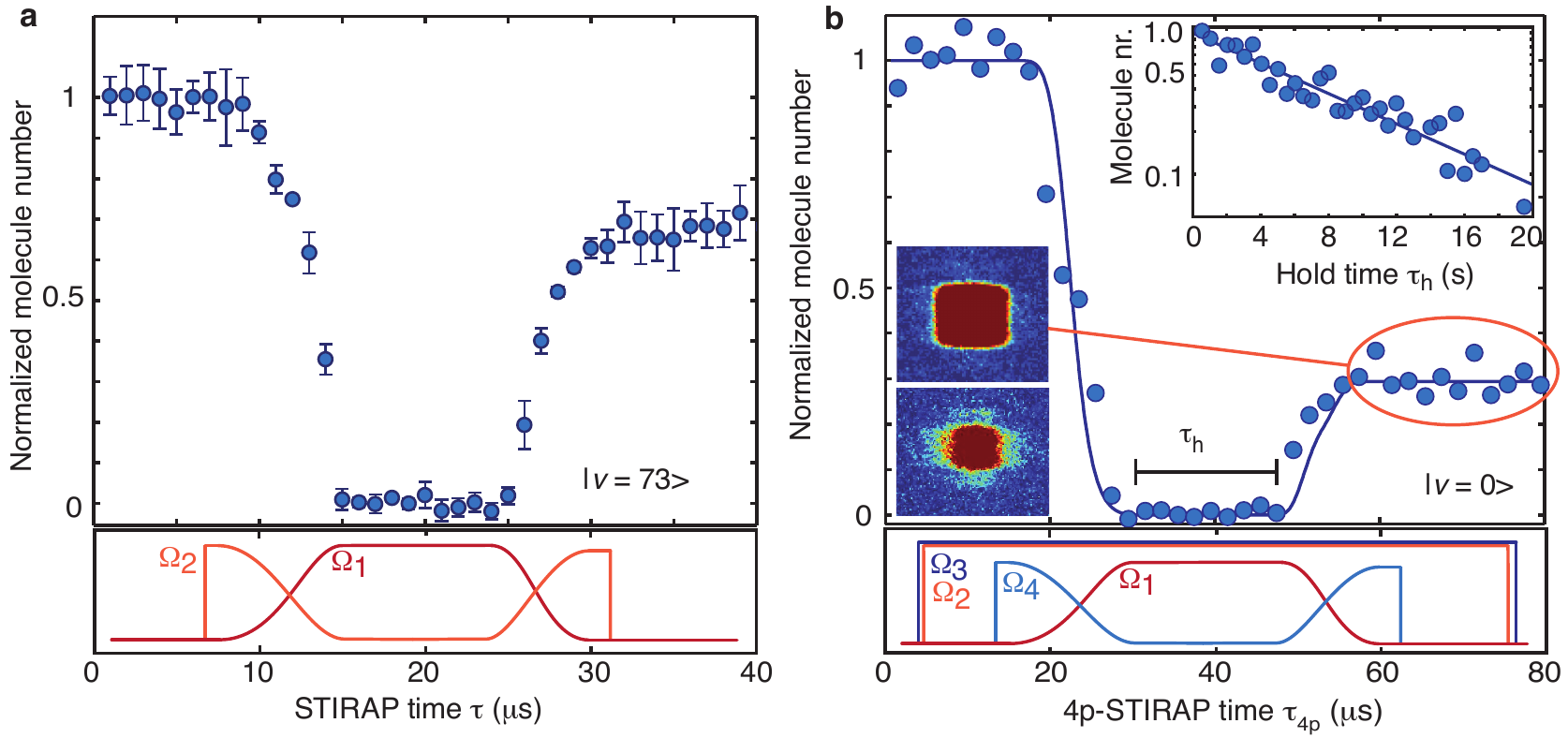}
\caption{\label{ICAPfig3}{\bf Transfer to the intermediate ground state level $|v=73\!\!>$ and to the rovibronic ground state $|v=0,J=0\!\!>$: a,} Transfer to $|v=73,J=2\!\!>$ and schematic pulse sequence for the Rabi frequencies $\Omega_i$. The number of Feshbach molecules in $|1\!\!>$ is monitored as a function of STIRAP time $\tau$ after which the transfer is interrupted. As the Rabi frequencies are ramped, the molecules are transferred to $|v=73\!\!>$ and no molecules are detected in $|1\!\!>$. For detection, the sequence is reversed and a large proportion of the molecules returns to $|1\!\!>$. Data taken from Ref. \cite{Danzl2008qgo}. {\bf b,} Transfer to the rovibronic ground state $X^1\Sigma_g^+ |v=0,J=0\!\!>$ by 4-photon STIRAP with schematic pulse sequence. The molecules are transferred in a single step to $|v=0\!\!>$ as the Rabi frequencies for lasers L$_1$ and L$_4$ are ramped while L$_2$ and L$_3$ continuously couple all intermediate levels. For detection, the pulse sequence is reversed. The solid line is a model calculation. Upper right inset: Normalized number of ground state molecules as a function of hold time $\tau_h$ in the lattice. STIRAP lasers were turned off during $\tau_h$ and the lattice depth was adiabatically reduced after transfer to $|v=0,J=0\!\!>$ to reduce off-resonant light scattering. Upper left inset: Absorption image after STIRAP to $|v=0,J=0\!\!>$ and back showing the momentum distribution after time of flight, corresponding to an average of the circled data points. Almost all the molecular population resides in the first Brillouin zone characterized by its rectangular shape. Lower left inset: Momentum distribution after transfer to $ |v=0,J=2\!\!>$ and back with resonant modulation of lattice depth during $\tau_h$, showing molecular population transferred to higher lattice bands. Data taken from Ref. \cite{Danzl2010auh}.}
\end{center}
\end{figure}

We first investigate transfer in the near quantum degenerate regime to the intermediate ground state level $|3>=|v=73,J=2\!\!>$ in the absence of an optical lattice \cite{Danzl2008qgo}. This presents the crucial transfer step because it links the long-range molecular levels near threshold with the tightly, i.e., chemically bound levels deep down in the molecular potential, bridging a binding energy difference of $\sim 1060$ cm$^{-1}$ and most of the mismatch in internuclear distance. As shown in Fig.\,\ref{ICAPfig3}a we monitor the number of Feshbach molecules as a function of STIRAP time $\tau$ during transfer to $|v=73,J=2\!\!>$, with the pulse sequence schematically depicted in the lower panel. After the STIRAP pulse sequence, no molecules remain in $|1\!\!>$. For detection of deeply bound molecules, they are transferred back to the original weakly bound level by a second, reverse pulse sequence. Molecules in $|1\!\!>$ are dissociated to atoms, which are then detected by standard absorption imaging. A large proportion of the molecules returns to $|1\!\!>$ after round-trip STIRAP, corresponding to a single pass transfer efficiency of $>$80\%. The coherent nature of the transfer was demonstrated in a Ramsey-type experiment and it was shown that the STIRAP process does not lead to heating of the sample on the nano-Kelvin scale \cite{Danzl2008qgo}.

Coherent state transfer in an optical lattice poses additional challenges that have to be met: (1) The lattice light might lead to excitation of molecules to electronically excited states and hence to loss out of the three- or five-level system. (2) The lattice light might create unwanted coherences competing with the STIRAP process, e.g., between an electronically excited intermediate level and an additional ground state level. (3) In order to ensure control over the motional state of the ground-state molecules in the optical lattice, the lattice has to be operated near the magic wavelength that gives equal light shifts for the initial and the final molecular levels. In order to realize lattice-based STIRAP to the rovibrational ground state, we first implemented transfer to $X |v=73,J=2\!\!>$ in the optical lattice \cite{Danzl2009dbu}. Transfer efficiencies similar to the case without an optical lattice were achieved. The lifetime of the molecular population in $|v=73\!\!>$ was on the order of 20 ms, limited by off-resonant excitation by the lattice light. This lifetime is orders of magnitude longer than the timescale of the STIRAP transfer, allowing us to use this deeply bound level as an intermediate state for lattice-based STIRAP transfer to $|v=0\!\!>$.

Figure \ref{ICAPfig3}b summarizes our central results on transfer to the rovibronic ground state in an optical lattice. We perform 4p-STIRAP and monitor the number of molecules in $|1\!\!>$ as a function of 4p-STIRAP time $\tau_{4p}$. Lasers L$_2$ and L$_3$ continuously couple all intermediate levels $|2\!\!>$, $|3\!\!>$, and $|4\!\!>$ to one effective intermediate level \cite{Vitanov1998apt} and STIRAP is carried out by varying only the Rabi frequencies $\Omega_1$ and $\Omega_4$ of the first and last laser. Population in the intermediate ground state level $|3\!\!>$ can in principle be minimized during transfer by choosing $\Omega_2, \Omega_3 \gg \Omega_1, \Omega_4$.
The molecules are directly transferred to $X^1\Sigma_g^+ |v=0,J=0\!\!>$. No molecules are detected during the hold time $\tau_h$.
The reverse pulse sequence transfers a large proportion of the molecules back to $|1\!\!>$, corresonding to a single pass transfer efficiency of $\sim$ 55 \%. We have achieved similar transfer efficiencies of 55 \% - 60 \% with s-STIRAP.

Remarkably, motional state control is maintained during STIRAP to $|v\!=\!0\!\!>$. During the STIRAP process, the wavefunction of the motional ground state in the lattice potential experienced by molecules in level $|1\!\!>$ is projected onto the motional wavefunctions associated with the lattice potential experienced by molecules in $|5\!\!>$. Only if the lattice wavelength is chosen such that the dynamical polarizabilities for $|1\!\!>$ and $|5\!\!>$ and hence the respective lattice potentials are closely matched will this projection lead to ground state molecules that are confined to the lowest vibrational level at the individual lattice sites.
The upper left inset in Fig.\,\ref{ICAPfig3}b shows an absorption image after STIRAP to $|v=0\!\!>$ and back after time of flight to reveal the momentum distribution of the molecules. The rectangular shape of the first Brillouin zone can clearly be seen. We deduce that 92(3) \% of the molecular population resides in the lowest lattice vibrational level. For comparison, the lower left inset in Fig.\,\ref{ICAPfig3}b shows molecular population transferred to higher lattice vibrational levels by amplitude modulation of the lattice during hold time $\tau_h$ with a frequency that matches the energy spacing between two lattice vibrational levels. By applying lattice amplitude and phase modulation and varying the modulation frequency we can map out the energy structure of the lattice bands for the $|v=0\!\!>$-molecules and deduce their dynamical polarizability. We find that the polarizability of the $|v=0\!\!>$-molecules is 1.05 $\times$ the polarizability of the Feshbach molecules \cite{Danzl2010auh}, such that the magic wavelength condition is well fulfilled.
We have thus created a state where the molecules are controlled at the level of single quantum states both with respect to all internal as well as the external degrees of freedom. In the central region of the lattice, about every other lattice site is filled with a molecule in the desired quantum state.

We measure the lifetime of the ground state molecules in the lattice by varying the hold time $\tau_h$ for up to 20 s. The right inset in Fig.\,\ref{ICAPfig3}b shows the number of rovibronic ground state molecules as a function of $\tau_h$. We observe a long  $1/e$ lifetime of 8.1(6) s, limited by off-resonant excitation to excited state molecular levels by the lattice light.

\begin{figure}[h]
\includegraphics[width=72mm]{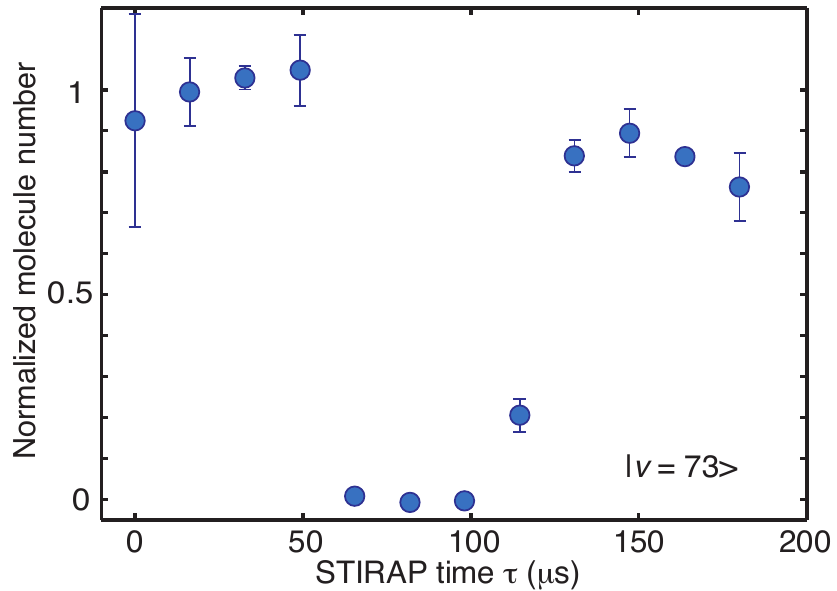}\hspace{2pc}
\begin{minipage}[b]{15pc}\caption{\label{ICAPfig4}{\bf Improved STIRAP efficiency:} Transfer to the intermediate ground state level $|3\!\!>=\!\!|v\!\!=\!\!73,J\!\!=\!\!2\!\!>$ with improved STIRAP laser setup. A substantial increase in transfer efficiency as compared to the data in Fig.\,\ref{ICAPfig3}a is achieved by increasing the laser phase stability.}
\end{minipage}
\end{figure}

Our simulations, based on the time dependent solution of the 5-level quantum master equation including laser phase noise and loss out of the 5-level system due to spontaneous emission, show that the STIRAP transfer efficiency is currently limited by finite available laser power to drive the extremely weak transitions in the 4p-scheme and by phase fluctuations of the STIRAP lasers. In order to optimize sample purity and hence phase-space density for quantum-state resolved collisional studies and the preparation of a BEC of ground state molecules, we have recently optimized both the short term phase coherence and the long term stability of our transfer lasers. For short term stability on the timescale of an individual STIRAP transfer process, a Pound-Drever-Hall lock with a feedback bandwidth of several MHz suppresses fast noise. In addition, we have modified the diode laser external grating feedback setup to essentially increase the laser resonator length to reduce phase noise. These improvements are combined with a slow phase lock of all STIRAP lasers to a common reference laser via the optical frequency comb. For this, we have implemented a transfer oscillator scheme, which eliminates the correlated noise contributions from the optical frequency comb and allows the transfer of the phase information of the reference laser to distinct regions of the optical spectrum \cite{Stenger2002umo,Telle2002klm}.
Testing the improved locking scheme on the fist STIRAP step to $|v\!\!=\!\!73,J\!\!=\!\!2\!\!>$, we have been able to improve the single pass efficiency by a substantial amount from $>$80\% to $>$90\%, as evidenced in Fig.\,\ref{ICAPfig4}. This indicates that the limitations to transfer efficiency are mainly technical. Additional technical improvements are expected to improve this efficiency further and a similar increase is expected for the second STIRAP transition.

\section{Conclusion}
We have shown that it is possible to prepare with high fidelity a sample of rovibronic ground-state molecules at ultralow temperatures in the presence of an optical lattice.
If the collisional properties of the ground state molecules turn out to be sufficiently favorable, the preparation of a BEC of ground state molecules should be possible. The most important obstacle to BEC formation would be an open channel for inelastic collisions in a dimer-dimer collision. However, it is known from ab-initio calculations that the channel for the formation of a trimer and an atom in a dimer-dimer collision is energetically closed for the alkali dimers \cite{Guerout2009cre,Zuchowski2010rou}.
We believe that our results can be generalized to the case of heteronuclear diatomic molecules such as RbCs. We have dedicated a separate experimental apparatus to this molecule \cite{Lercher2010poa}. It has the advantage that it is chemically stable under a two-body collision, i.e., the reaction RbCs + RbCs $\to$ Rb$_2$ + Cs$_2$ is endothermic when RbCs is in its rovibronic ground state \cite{Zuchowski2010rou}. In contrast, the molecule KRb is not stable \cite{Ospelkaus2010qsc} under the process KRb + KRb $\to$ K$_2$ + Rb$_2$. In principle, we expect that at ultralow temperatures and in the lowest hyperfine sublevel of the rovibronic ground state of the RbCs molecule \cite{Aldegunde2008hel} only molecular three-body collisions present a possible limitation to performing experiments with dipolar many-body systems. As before, to optimize the overall molecule production efficiency by minimizing collisional loss during the preparation procedure, we will implement the Feshbach association process and the ground state transfer in the presence of the optical lattice with precisely one molecule per lattice site. Evidently, the preparation of the initial two-atom Mott-insulator states is more involved than for the case of Cs$_2$ \cite{Damski2003coa,Freericks2010ite}. The central question is how to generate a many-body state in the lattice for which the number of Rb-Cs pairs as precursors to RbCs molecules is maximized. In view of considerable Cs-Cs and in particular Rb-Cs three-body loss \cite{Lercher2010poa} we will load the lattice with two spatially separated BECs, one BEC of $^{87}$Rb atoms and one BEC of $^{133}$Cs atoms. The two BECs initially reside in two separate dipole traps. We will exploit the fact that the Rb-Cs interaction can be controlled near a Rb-Cs interspecies Feshbach resonance \cite{Pilch2009ooi}. The central idea is to first freeze out the Cs sample using the SF-MI phase transition in such a way that at first a state with precisely one Cs atom per lattice site is generated. The Rb sample, much more mobile as a result of a lower polarizability for the Rb atoms, is then brought to overlap with the Cs sample while being still superfluid. For small repulsive or even weakly attractive Rb-Cs interaction one expects that each Cs atom will then preferentially pair up with precisely one Rb atom, giving ideal conditions for molecule production and ground-state transfer.

\section{Acknowledgments}
We thank H. Ritsch, N. Bouloufa, O. Dulieu, J. Aldegunde, J.~M. Hutson, H. Salami, T. Bergeman, S. D\"urr, and K. Bergmann for valuable discussions and H. Telle, H. Schnatz, B. Lipphardt, and J. Alnis for sharing technical expertise. We are indebted to R. Grimm for generous support. We gratefully acknowledge funding by the Austrian Ministry of Science and Research (Bundesministerium f\"ur Wissenschaft und Forschung) and the Austrian Science Fund (Fonds zur F\"orderung der wissenschaftlichen Forschung) in the form of a START prize grant and by the European Science Foundation within the framework of the EuroQUASAR collective research project QuDeGPM (Project  I 153-N16) and within the framework of the EuroQUAM collective research project QuDipMol (Project I 124-N16). R.H. was supported by a Marie Curie International Incoming Fellowship within the 7th European Community Framework Programme.

\section*{References}
\medskip
\providecommand{\newblock}{}

\smallskip

\end{document}